\documentclass[prl,aps,twocolumn,showpacs]{revtex4}
\usepackage{graphicx}

\begin{document}
\title{On Spin-Glass Complexity}

\author{A. Crisanti, L. Leuzzi, G. Parisi and T. Rizzo}
\affiliation{Dipartimento di Fisica, SMC and INFM, Universit\`{a} di
Roma ``La Sapienza'', 
P.le A. Moro 2, I-00185 Roma, Italy}

\begin{abstract}
We study the quenched complexity in spin-glass mean-field models
satisfying the Becchi-Rouet-Stora-Tyutin supersymmetry.
The outcome of such study, consistent with recent numerical results,
allows, in principle, to conjecture the absence of {\em any
supersymmetric contribution to the} complexity  in the
Sherrington-Kirkpatrick model. 
The same analysis can be applied to any model with 
a Full Replica Symmetry Breaking phase, e.g. the Ising
$p$-spin model below the Gardner  temperature. 
The existence of different solutions, breaking the supersymmetry,
 is also discussed. 
\end{abstract} 


\maketitle 

Mean field spin glass models can display different kinds of {\em
frozen phase} according to the choice of the interaction between their
constituent elements. Such a phase can be either Full Replica Symmetry
Breaking (FRSB), i.e. described by means of an order parameter that is
a {\em function} of the replica-group index $x$: the {\em overlap}
$q(x)$, or One step Replica Symmetry Breaking (1RSB), in which only
two groups of replicas are necessary for a proper representation of
the properties of the system and the overlap consequently takes only
two values.  The prototype of the first kind is the
Sherrington-Kirkpatrick (SK) model \cite{SKPRL75}, whereas for the
second kind we will mainly discuss the Ising $p$-spin model,
displaying discontinuous overlap at the paramagnet/spin-glass
transition at temperature $T_s$ and, moreover, a second 1RSB/FRSB
transition deep in the frozen phase, at the Gardner temperature $T_G$
\cite{Gard}.

Decreasing the temperature from the paramagnetic phase, both the SK
and the $p$-spin models undergo a transition to a phase where a
dynamic aging regime sets up. Below the transition temperature the
infinite system never equilibrates: dynamical two-time quantities such
as correlation and response function are not time translational
invariant and do not satisfy the fluctuation-dissipation theorem.
However, the nature of the aging regime is quite different in the two
models. Furthermore, in the SK model such transition coincides with
the static transition, taking place at temperature $T_s$, and, at
finite temperature, the large time values of one-time intensive
quantities, e.g.  the energy, tend to their equilibrium value. In
$p$-spin models, instead, the temperature at which the aging regime
arises, called $T_d$, is above $T_s$ and the dynamical energy never
converges to its equilibrium value, remaining above some threshold
$E_{\rm th}$.  It is largely believed that the dynamical features of
these systems are connected to the presence of a high number of
metastable states, { the logarithm of which, divided by the size of
the system, is called Configurational Entropy or Complexity.}

In this letter we present {a quenched FRSB computation of the
complexity of the SK model based on the Becchi-Rouet-Stora-Tyutin
supersymmetry (BRST-susy).  While in 1RSB models the complexity saddle
point is supersymmetric and finite, in the SK model our computation
shows that a SUSY solution would be compatible only with a
subextensive number of metastable states}.  It is reasonable to think
that some of the differences between the dynamics of the two classes
of models amount to this difference in behavior of the complexity.

We will now concentrate on the complexity of the SK model, later
discussing what insight we can gain from the study of the Ising
$p$-spin model.  The SK model consists of $N$ Ising spins connected to
each other by random variables $J_{ik}$ of variance $1/N$. The states
of the system are usually identified with a proper subset of solutions of the
Thouless-Anderson-Palmer (TAP) \cite{TAP} equations, the computation
attempts to count these solutions.  The TAP equations are mean-field
equations for the single-site magnetizations $\{m_i\}$:
\begin{equation}
m_i=\tanh \left[\beta \left(\sum_{k=1}^N J_{ik}m_k-m_i (1-q)\right)\right]
\end{equation}
Where $q=\sum_i m_i^2/N$.  They can be obtained extremizing the
following TAP free energy function $F(\{m_i\})$ with respect to the
$\{m_i\}$:
\begin{eqnarray}
&&\beta F(\{m\}) = 
-\beta \sum_{i<j}J_{ij} m_i m_j
-\frac{N \beta^2}{4}    \left(1-q\right)^2
\\
&&
\nonumber
+\sum_i\left\{ {1 +m_i \over 2}\ln {1 +m_i \over 2}+ {1 -m_i \over
2}\ln {1 -m_i \over 2} \right\}
\end{eqnarray}

Not all TAP solutions can be
identified with physical states. An important condition is that
they must be minima of the TAP free energy. Furthermore, in order to
yield the correct magnetic susceptibility of a state in zero external
field, i.e. $\chi=\beta (1-q)$, the solutions must satisfy the {\em
Plefka criterion} $x_p \equiv 1-\beta^2 \sum (1-m_i^2)^2/N \geq
0$ \cite{Ple1}.

This criterion is encountered also in the replica computation of the
equilibrium free energy as the central stability condition of the
saddle point with respect to fluctuation of the order parameter
$Q_{ab}$, indeed it corresponds to the condition of positivity of the
replicon eigenvalue. Equivalently it can be recovered in the context
of the cavity method as the condition of positivity of the spin-glass
susceptibility $\chi_{SG}=\sum_{ij} \chi_{ij}^2/N$ \cite{MPV}.

We would like to compute the minima of the TAP free energy functional
satisfying the Plefka criterion. Unfortunately the only expression one
can handle is:
\begin{equation}
\rho_s(f)=\int_{-1}^{1}\prod_{i=1}^N d m_i~ \delta\left(\partial_{i}
F\right) {\rm det}(\partial_{i}\partial_{j}F) \delta\left(F - N
f\right)
\end{equation}
Strictly speaking this corresponds to count all the solutions of the
TAP equations of a given free energy $f$ weighted with the sign of the
determinant of the Hessian.  Therefore the assumption that this
expression corresponds to the number of physical states must be
justified somehow {\em a posteriori}. Through standard manipulations
we can express $\rho_s$ as an   integral over fermionic, 
$\{\psi_i,\overline{\psi}_i\}$, and  bosonic, $\{ m_i,x_i\}$, 
variables   of the exponential of the
 following action \cite{K91,CGPM}:
\begin{equation}
{\cal S}=
x_i \partial_{m_i} F(\{m\}) 
+{\overline\psi}_i\psi_j\partial_{i}\partial_{j}F(\{m\})
+u(F(\{m\})-N f)
\end{equation}
The variable $u$ is given by the $\delta$-function over the free energy.
This action posses the 
BRST-susy \cite{CGPM}, indeed it is invariant under the transformation
 $\delta m_i=\epsilon ~\psi_i \ ; \ \delta x_i=\epsilon~ u~\psi_i\ ;\
 \delta{\overline\psi}_i=-\epsilon~x_i \ ; \ \delta \psi_i = 0$.  We
 can perform annealed or quenched averages yielding respectively $\ln
 \overline{\rho_s}$ and $\overline{\ln \rho_s}$; the quenched are the
 physical ones, i.e. those describing the properties of a typical
 system. They can be computed through the replica method, considering
 $n$ copies of the system and taking the limit $n \rightarrow 0$.
 Eventually one obtains an integral of the exponential of a
 'macroscopic' action ${\cal S}_{\rm macro}$ depending on four bosonic
 and four fermionic (in case replicated) variables (see
 e.g. \cite{K91}).  If the complexity is extensive the integral may be
 evaluated by steepest descent, solving saddle-point equations for the
 macroscopic action. Furthermore, the fermionic part always gives a
 subextensive contribution and can be neglected, yielding the well
 known expression obtained more than twenty years ago by Bray and
 Moore (BM) \cite{BMan,BMque, CGPM, noian}.

The problem of finding a solution to the saddle-point equations
obtained in order to extremize the macroscopic action is highly non
trivial. The annealed case, widely studied in the literature
\cite{BMan,CGPM,noian}, tells us that there are different solutions to
these equations and we face the problem of selecting, if any, the
correct one.  In particular, there is a BRST-susy solution and a
BRST-susy-breaking solution.  The selection problem has been recently
considered by the authors \cite{noian} and it has been shown that all
solutions currently known for the annealed case present some problems.
In the quenched case it has been shown \cite{BMY} that the complexity
curve $\Sigma (f)$, if it exists, vanishes at the equilibrium free
energy.  In \cite{noian} it has been pointed out that such point of
the complexity curve is described by a BRST-susy saddle point.

The fermionic, subextensive part, usually neglected in the computation
 of the complexity by means of the saddle point approximation, is
 actually very important: taking it into account one sees that the
 macroscopic action too is invariant under a proper macroscopic
 BRST-susy transformation between its eight (eventually replicated)
 variables \cite{K91} { and this causes} the prefactor of the
 exponential of the BM saddle point to be zero at all orders in an
 expansion in $1/N$.

We have studied  the quenched complexity satisfying the BRST-susy.
 In \cite{BMan} the existence of a
replicated BRST-susy-breaking solution is hypothesized, with a complexity
which goes to zero at the equilibrium free energy. This solution, however,
has not been exhibited up to now. We will come back to this point
below.

Considering a BRST-susy saddle point, considerably simplifies the
 computation of $\int e^{{\cal S}_{\rm macro}}$. Indeed, as already
 discussed in \cite{CGPM, noian}, one recovers the computation scheme
 proposed in \cite{MPRL95,PP}. We, thus, consider the
 replicated expression of the variational free energy used to compute
 the equilibrium free energy, which depends on the $n \times n$ matrix
 $Q_{ab}$. In the limit $n \rightarrow 0$, $Q_{ab}$ is parametrized by
 means of the Parisi Ansatz in terms of the function $q(x)$ defined in
 the interval $[0 ,1]$. 
 We impose by hand a break point of $q(x)$ at $x=m<x_{\rm
 static}$ (where $x_{\rm
 static}$ is the break point of the Parisi solution)
and then we extremize the free energy
 with respect to $q(x)$ obtaining the function $F(\beta,m )$. 
The  complexity $\Sigma (\beta,f)$ at a given free energy $f$ is obtained
through the following equations:
\begin{equation}
\Sigma (\beta,f)=\beta m^2 {\partial F  \over \partial m}\ ,\ 
 f={\partial (m F(\beta,m))  \over  \partial m}
\label{eqmon}
\end{equation}
with 
\begin{equation}
F(\beta,m) =\mbox{ext} -{\beta \over 4} \left[1-2q(1)
+\int_0^1q^2(x)dx \right]-f(0,0)
\label{Fm}
\end{equation}
The function $f(x,y)$ satisfies the Parisi differential equation
$\partial_x f = \dot{q} / 2 [\partial^2_y f 
+x\left(\partial_y f  \right)^2]$
with initial condition:
\begin{equation}
f(m^-,y)={1 \over \beta m}\ln \int dp_{\Delta q}(z) \cosh^m (\beta
y+\beta z)dz+{\ln 2 \over \beta }\
\end{equation}
where $dp_{\Delta q}(z) $ is a Gaussian measure over $z$ with zero
mean and variance $\Delta q$ which accounts for a possible
discontinuity at $x=m$.
We have computed $q(x)$ as a function of $m$ and the resulting
complexity both by numerical integration and by an exact series expansion
in powers of the reduced temperature through the techniques of
\cite{Techn}.  
We find that setting $m$ smaller than the
static break point $x_{\rm static}$ the function $q(x)$ which
extremizes the free energy develops a discontinuity at $x=m$.  In
general, $q(x)$ is a continuous monotonous function for $x<m$,
 at some $x_0 < m$ it develops a plateau such that
$q(x)=q(x_0)$ for $x_0 \leq x \leq m$: in particular $q(m^-)=q(x_0)$;
it displays, then, a positive discontinuity $\Delta q$ at $x=m$, while it is
constant for $x>m$.

\begin{figure}[b!]
\begin{center}
\includegraphics*[width=.49 \textwidth, height=.24 \textheight]{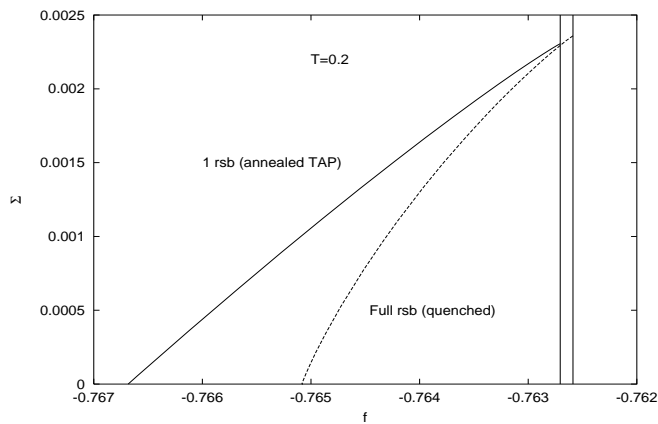}
\caption{Annealed and quenched complexity of the BRST solution in SK model.}
\label{figure3}
\end{center}
\end{figure}

At some $m=m_{{\rm max}}$, corresponding to a threshold free energy
$f_{\rm th}$, $\Sigma(\beta,m)$ takes its maximal value (see figure
\ref{figure3}).  We can identify $\Sigma (f_{\rm th})$ with the total
complexity.

Our most important result is that this solution must be rejected on
physical ground.  In figure \ref{figure3} the BRST-susy quenched
complexity is plotted versus $f$ together with the BRST
annealed complexity \cite{CGPM, noian}. Convexity implies that $\ln
\overline{\rho_s}\geq \overline {\ln \rho_s} $. The total
complexities, in the annealed and quenched case, can be identified
with those of the states with higher complexity, i.e. $\Sigma_{\rm
tot}=\Sigma(f_{\rm th})$ and, thus,  the previous convexity
condition is violated. Moreover, convexity implies that at any $f$ the
annealed complexity must be greater then the quenched one, but $f_{\rm
th}$ is greater in the quenched case, therefore there is a region
where the annealed complexity is zero while the quenched one is finite
thus violating this condition.

Most importantly, we have checked (details elsewhere \cite{CLPR3})
that the solution does not satisfy the Plefka criterion, {\it i.e.}
the replicon eigenvalue is negative as soon as $m<x_{\rm static}$.
This means that, provided it actually describes some TAP solutions,
they have no physical meaning. As discussed in \cite{noian} the
violation of the Plefka criterion leads also to a mathematical
inconsistency with some assumptions implicit in the solution (i.e. the
condition $B=0$, see \cite{noian}). A direct computation shows that
this result can be extended to generic mean-field models with a
continuous FRSB $q(x)$.

Since the solution is unphysical { for any $f>f_{\rm eq}$, we
obtain that BRST-susy in the SK model implies zero complexity.  In
order to account for a finite complexity one has to look for another
solution (see discussion at the end of the letter).  In the BRST-susy
case} the number of states is subextensive and the states have all
the same free energy per spin, equal to the equilibrium value.
Furthermore all the states verify $x_p=0$. { We notice that}
all these properties are in complete agreement with both old and
recent numerical findings \cite{Ple2,BMless}{ , not implying
any supersymmetry.} In \cite{Ple2} the minima of the TAP free energy
with $x_p \geq 0$ (i.e. those that can be physically identified with
states of the system) are studied by means of some modified TAP
equations \cite{Ple2} which allow to separate this set from the
non-physical solutions, in \cite{BMless} the original equations are
employed.  In both cases turns out that all these minima satisfy
$x_p=0$, as $N\to \infty$, in disagreement with the prediction of the
{ BRST-susy-breaking solution \cite{BMY,noian}.}
{ We also mention that a zero complexity in the SK model
was also obtained in the dynamical reformulation of Parisi solution
performed in Ref.~\cite{vMCJPA01}.
}
\begin{figure}[b!]
\begin{center}
\includegraphics*[width=.49 \textwidth, height=.24 \textheight]{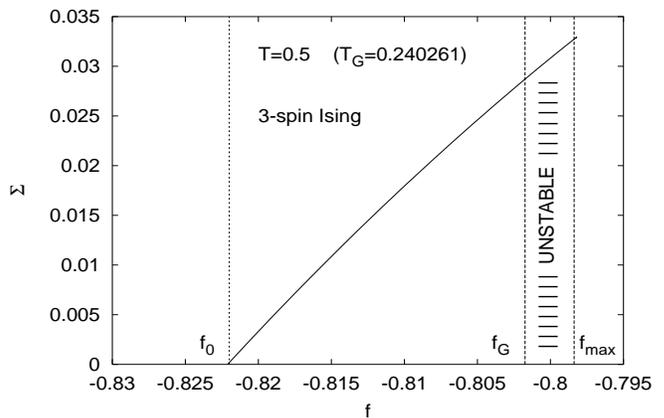}
\caption{Complexity of the Ising $p$-spin model, for $p=3$.
It counts (meta-)stable states up to a value
$f_G$, lower than $f_{\rm max}$ 
(it would be $f_G=f_{\rm max}$  in the
spherical $p$-spin case 
where no 1RSB/FRSB transition occurs).  
Beyond such point the stability condition (Plefka's criterion)
is violated.
In particular, $f_G\rightarrow f_0$
as $T \rightarrow T_G$.}
\label{figure1}
\end{center}
\end{figure}

The Ising $p$-spin model \cite{Gard} is useful to understand the
different behavior of the complexity in 1RSB and FRSB models. This
system undergoes a first phase transition from a paramagnetic to a
1RSB phase at a temperature $T_s$ and a second, continuous, phase
transition from the 1RSB phase to a FRSB phase at $T_G<T_s$. The
replicon eigenvalue of the 1RSB solution is positive in the 1RSB phase
but goes to zero at $T_G$. At $T<T_G$ the replicon is negative on the
1RSB solution while it is strictly zero on the 
FRSB solution.
At temperatures $T_G<T<T_s$ we can compute the quenched complexity
using the standard recipe of \cite{MPRL95}
using a
1RSB order parameter. The result, see e.g. figure 2, is a quenched
complexity which is zero at the equilibrium free energy and reaches a
maximum value at some $f_{\rm max}$. 
However, as noted in \cite{RM}, the replicon eigenvalue,
which is positive at $f=f_{0}$, goes to zero at some $f_G<f_{\rm max}$
and is negative at free energies $f>f_G$ where the solution must be
rejected, see figure \ref{figure1}. Thus we identify $f_G$ as the free
energy threshold at that temperature { if we stick to the SUSY
solution.} It turns out that $\lim_{T\rightarrow T_G}
f_G(T)=f_{0}(T_G)$. Therefore, approaching $T_G$ from above, the range
$[f_{0},f_G]$ where the complexity is finite shrinks to zero.
Equivalently, the total complexity goes to zero at $T_G$:
$\lim_{T\rightarrow T_G} \Sigma(T)=\lim_{T\rightarrow T_G}
\Sigma(f_G(T))=\Sigma (f_{\rm eq}(T_G))=0$. Therefore, at the
transition from a 1RSB equilibrium phase to a FRSB equilibrium phase
the complexity vanishes and remains zero in the whole FRSB phase.

{
We now discuss
the possible existence of  other 
quenched solutions, such as the one hypothesized in \cite{BMY}.
}
BRST-susy-breaking solutions certainly exist in the annealed case
while they are not known in the quenched case. However, as discussed
in Refs. \cite{K91,noian} they would need some justification in
order to be used.  Taking into account corrections to the saddle point
it turns out that the expansion in powers of $1/N$ of the prefactor of
the leading exponential contribution is zero at all orders.  This
result has been obtained by Kurchan \cite{K91} for the total
complexity of the naive TAP equations exploiting the BRST-susy of the
macroscopic action referred above; we have extended this result to the
SK model at a generic value of $u$ (the variable conjugated to the
free energy), i.e. at any value of the free energy. Clearly a zero or
exponentially small prefactor will dramatically change the saddle
point prediction.  Furthermore, according to \cite{BMan} the quenched
BRST-susy-breaking solution should coincide with the annealed
BRST-susy-breaking solution at free energies greater than some $f_c$
and it has been shown \cite{noian} that this solution does not compare
well with the numerical data of \cite{Ple2,BMless}.  In particular, it
predicts a finite value of $x_p$ which is out of the error bars of the
numerical prediction, which instead hints a zero value.
Discrepancy are also found in the support of the complexity
\cite{Ple2}. 
 We also mention that in the
Ising $p$-spin model at $T_G<T<T_s$ a BRST-susy-breaking solution
similar to the annealed one of the SK model can be found
\cite{Rieger}.  This solution, however, yields a complexity vanishing
at a free energy different from the equilibrium one at a given
temperature and violating Plefka's criterion for low free energies (see
\cite{CLPR3} for details). On the contrary, the BRST-susy solution of
the Ising $p$-spin model vanishes at exactly the equilibrium 1RSB free
energy (Fig.~\ref{figure1}).

{{ Quite recently it has been shown that the TAP solutions counted
by the BM complexity, besides a strictly positive spectrum display an
isolated eigenvalue going to zero for $N\to \infty$\cite{ABM}. This
property solves the apparent violation of the Morse theorem
\cite{CGPM}. In general, this result shows a substantial
difference between the BRST-susy and BRST-susy-breaking solution: only
the first one describes TAP solutions with a strictly positive
spectrum. 
Therefore, together with our results, this { leads} to the
conclusion that, while in 1RSB systems  a BRST-susy complexity
counts an extensive number of states with strictly positive
eigenvalues, in FRSB systems the BRST-susy solution is unphysical and the
BRST-susy-breaking solution proposed by BM counts an extensive number of
configurations with a flat direction out, as $N\to\infty$.}}

Regarding the possible existence of quenched BRST-susy solutions other than
the one we considered,  they would violate Plefka
criterion too, yielding a negative replicon eigenvalue.  Indeed, a
direct computation (details elsewhere \cite{CLPR3}) shows that in FRSB
models the violation of the criterion is caused by the discontinuity
at $x=m$, while a continuous function 
marginally satisfies it. 
On the other hand, if $q(x)$
is continuous at $x=m$, it holds $\partial ( F(m, \beta)) / \partial m =0$ 
and we will not be able to apply the recipe of Eq. \ref{eqmon}.
The discontinuity at $x=m$ is therefore unavoidable
in order to have a non trivial complexity but, at the same time, it
causes instability.

Looking at the Ising  $p$-spin model we recall that in \cite{RM} the
existence of states with free energy higher than $f_G$ has been
hypothesized: since the 1RSB solution is unstable, the complexity of
these states should be described by a FRSB solution, much as the
equilibrium free energy at $T<T_G$. It is not known whether such a
solution actually exists but it can be proved that  violation of marginal 
stability will occur in the $p$-spin model as well: 
$q(x)$ should be discontinuous at $x=m$ but at the same time
the presence of a FRSB region on the left of the discontinuity will
force the replicon to be negative. The problem of the
existence of a complexity of the clusters \cite{RM} in the $p$-spin
model remains, instead, open.

\end{document}